\documentstyle{jfm}
\input psfig.sty

\title[Heat transport by turbulent Rayleigh-B\'enard Convection]{Heat transport by turbulent Rayleigh-B\'enard Convection in cylindrical samples with aspect ratio one and larger} 
\author{Denis Funfschilling, Eric Brown, Alexei Nikolaenko, and Guenter Ahlers}
\affiliation{Department of Physics and iQUEST,\\ University of
California, Santa Barbara, CA  93106}
\begin{document}

\maketitle

\begin{abstract}
We present high-precision measurements of the Nusselt number $\cal N$ as a function of the Rayleigh number $R$ for cylindrical samples of water (Prandtl number $\sigma = 4.38$) with diameters $D = 49.7, 24.8,$ and 9.2 cm, all with aspect ratio $\Gamma \equiv D/L \simeq 1$ ($L$ is the sample height). In addition, we present data for  $D = 49.7$ and $\Gamma = 1.5, 2, 3,$ and 6. For each sample the data cover a range of a little over a decade of $R$. For $\Gamma \simeq 1$ they jointly span the range  $10^7 \stackrel {<}{_\sim} R \stackrel {<}{_\sim} 10^{11}$. Where needed, the data were corrected for the influence of the finite conductivity of the top and bottom plates and of the side walls on the heat transport in the fluid to obtain estimates of $\cal N_{\infty}$ for plates with infinite conductivity and sidewalls of zero conductivity. For $\Gamma \simeq 1$ the effective exponent $\gamma_{eff}$ of ${\cal N}_{\infty} = N_0 R^{\gamma_{eff}}$ ranges from 0.28 near $R = 10^8$ to 0.333 near $R \simeq 7\times10^{10}$. For $R \stackrel {<}{_\sim} 10^{10}$ the results are consistent with the Grossmann-Lohse model. For larger $R$, where the data indicate that  ${\cal N}_\infty(R) \sim R^{1/3}$, the theory has a smaller $\gamma_{eff}$ than $1/3$ and falls below the data. The data for $\Gamma > 1$ are only a few percent smaller than the $\Gamma = 1$ results.

\end{abstract}

\section{Introduction}
\label{sec:introduction}

A central prediction of theoretical models of turbulent Rayleigh-B\'enard convection (RBC) in a fluid heated from below [\cite{Kr62,Si94,Ka01,AGL02,GL00}] is the dependence of the global heat transport on the Rayleigh number
\begin{equation}
R = \alpha g \Delta T L^3/\kappa \nu
\end{equation}
($\alpha$ is the isobaric thermal expension coefficient, $\kappa$ the thermal diffusivity,  $\nu$ the kinematic viscosity, $g$ the acceleration of gravity, $\Delta T$ the temperature difference, and $L$ the sample  height) and the Prandtl number $\sigma = \nu/\kappa$. The heat transport is usually expressed in terms of the Nusselt number
\begin{equation}
{\cal N} = Q L / \lambda \Delta T
\label{eq:nusselt}
\end{equation}
were $Q$ is the heat-current density and $\lambda$ is the thermal conductivity of the fluid in the absence of convection. Before a quantitative comparison between theory and experiment can be made, the results for $\cal N$ usually must be corrected for the influence of the side wall [\cite{Ah00,RCCHS01,NS03}] and the top and bottom plates [\cite{CCC02,Ve04,BNFA05}] to yield an estimate of the idealized ${\cal N}_\infty$.
 
 A model developed recently by \cite{GL00}, based on the decomposition of the 
kinetic and the thermal dissipation into boundary-layer and bulk contributions, provided a good fit to experimental data [\cite{XBA00}, \cite{AX01}] for a cylindrical sample of aspect ratio $\Gamma \equiv D/L = 1$ ($D$ is the diameter) when it was adapted [\cite{GL01}, GL] to the relatively small Reynolds numbers of the measurements. However, the data were used to determine four adjustable parameters of the model. Thus more stringent tests using measurements for the same $\Gamma$ but over wider ranges of $R$ and $\sigma$ are desirable. A success of the model was the agreement with recent results by \cite{XLZ02} for much larger Prandtl numbers than those of \cite{AX01}, for  $R = 1.78\times 10^7$ and $1.78\times 10^9$. It is the primary aim of the present paper to extend the range of $R$ over which  high-precision data, subject to minimal systematic errors, are available for ${\cal N}_\infty(R)$. Our data span the range  $10^7 \stackrel {<}{_\sim} R \stackrel {<}{_\sim} 10^{11}$ with $\sigma = 4.38$ and $\Gamma \simeq 1$ and deviate from the Boussinesq approximation (\cite{Bo1903}) by less than a few tenths of a percent. We believe that they can  serve as a benchmark for comparison with future experimental and theoretical developments. They agree quite well with the GL model for $R \stackrel {<}{_\sim} 10^{10}$, but for larger $R$ there are deviations. 

In addition to the results for $\Gamma \simeq 1$ we present also some data for larger $\Gamma$, up to $\Gamma = 6$. We find that there is remarkably little dependence of ${\cal N}$ on $\Gamma$. For instance, the $\Gamma = 6$ data fall only about 4\% below the $\Gamma = 1$ results.

\begin{table}
\begin{center}
\begin{tabular}{cccccccccccccc}
No& $\bar T (^\circ C)$ & $\Delta T (^\circ C)$ & $10^{-8}R$  & $\cal N$ & ${\cal N}_\infty$ & & No& $\bar T (^\circ C)$ & $\Delta T (^\circ C)$ &  $10^{-8}R$  & $\cal N$ & ${\cal N}_\infty$ \\
 1 & 40.009 &  1.957 & 94.3 & 127.0 & 129.3 & &  2 & 40.011 &  3.911 & 188.6 & 157.5 & 161.4\\
 3 & 39.984 &  5.917 & 285.0 & 179.4 & 184.8 & &  4 & 40.007 &  7.821 & 377.0 & 195.8 & 202.5\\
 5 & 40.007 &  9.764 & 470.7 & 210.1 & 218.0 & &  6 & 40.022 & 11.676 & 563.1 & 222.3 & 231.3\\
 7 & 40.039 & 13.589 & 655.7 & 233.5 & 243.6 & &  8 & 39.955 & 15.688 & 754.8 & 243.7 & 254.9\\
 9 & 39.901 & 17.729 & 851.4 & 253.4 & 265.6 & &  10 & 39.887 & 19.705 & 945.8 & 261.8 & 274.9\\
 11 & 40.041 &  6.783 & 327.3 & 187.5 & 193.5 & &  12 & 40.062 &  4.791 & 231.4 & 167.9 & 172.5\\
 13 & 40.056 &  2.849 & 137.6 & 142.8 & 145.9 & &  14 & 39.963 &  2.543 & 122.4 & 137.5 & 140.3\\
 15 & 39.944 &  1.595 & 76.7 & 118.5 & 120.4 & &  16 & 39.923 & 19.623 & 943.0 & 261.7 & 274.7\\
 17 & 39.921 & 19.627 & 943.2 & 261.6 & 274.7 & &  18 & 39.929 &  5.048 & 242.7 & 170.6 & 175.4\\
 19 & 39.970 &  1.050 & 50.6 & 104.6 & 106.0 & &  20 & 39.999 &  9.775 & 471.1 & 210.3 & 218.2\\
 21 & 39.998 &  9.782 & 471.4 & 210.3 & 218.3 & &  22 & 40.016 &  0.962 & 46.4 & 101.8 & 103.1\\
 23 & 40.015 &  0.963 & 46.4 & 101.9 & 103.2 & &  24 & 39.904 & 19.666 & 944.5 & 261.7 & 274.8\\
 25 & 39.963 &  2.539 & 122.2 & 137.6 & 140.4 & &  26 & 40.000 &  1.485 & 71.5 & 116.3 & 118.1\\
 27 & 40.011 &  1.954 & 94.2 & 127.0 & 129.2 & &  28 & 40.011 &  1.955 & 94.3 & 126.9 & 129.2\\
 29 & 40.010 &  1.954 & 94.2 & 126.9 & 129.1 & &  30 & 39.993 &  1.005 & 48.4 & 103.1 & 104.4\\
 31 & 39.859 & 21.687 & 1040.0 & 269.4 & 283.3 & &  32 & 39.971 &  3.988 & 192.0 & 158.4 & 162.4\\
\end{tabular}
\caption{Results for $\Gamma = 0.982$, run 2 from the large apparatus ($D = 49.7$ cm). In Table~\ref{tab:1.0l} to \ref{tab:6.0l} two points are listed per line, and they are numbered in chronological sequence.}
\label{tab:1.0l}
\end{center}
\end{table}

\begin{table}
\begin{center}
\begin{tabular}{cccccccccccccc}
No& $\bar T (^\circ C)$ & $\Delta T (^\circ C)$ & $10^{-8}R$  & $\cal N$ & ${\cal N}_\infty$ & & No& $\bar T (^\circ C)$ & $\Delta T (^\circ C)$ &  $10^{-8}R$  & $\cal N$ & ${\cal N}_\infty$ \\
 1 & 39.985 &  2.002 & 11.3 & 66.5 & 66.6 & &  2 & 40.014 &  2.437 & 13.7 & 70.5 & 70.6\\
 3 & 40.003 &  3.147 & 17.7 & 76.0 & 76.2 & &  4 & 39.968 &  4.005 & 22.6 & 81.7 & 81.9\\
 5 & 39.987 &  4.951 & 27.9 & 87.2 & 87.5 & &  6 & 39.973 &  6.202 & 34.9 & 93.3 & 93.7\\
 7 & 39.994 &  7.679 & 43.3 & 99.6 & 100.1 & &  8 & 39.946 &  9.731 & 54.8 & 107.0 & 107.6\\
 9 & 39.956 & 11.862 & 66.8 & 113.7 & 114.5 & &  10 & 39.928 & 14.259 & 80.2 & 120.3 & 121.2\\
 11 & 39.911 & 16.824 & 94.6 & 126.5 & 127.6 & &  12 & 39.865 & 19.836 & 111.3 & 133.1 & 134.5\\
 13 & 39.979 &  1.618 &  9.1 & 62.4 & 62.5 & &  14 & 39.998 &  1.282 &  7.2 & 58.3 & 58.3\\
 15 & 39.970 &  1.041 &  5.9 & 54.8 & 54.9 & &  16 & 39.968 &  0.845 &  4.8 & 51.6 & 51.6\\
 17 & 39.967 &  0.650 &  3.7 & 47.7 & 47.7 & &  18 & 39.989 &  0.507 &  2.9 & 44.4 & 44.5\\
 19 & 39.954 & 22.581 & 127.1 & 138.7 & 140.2 & &  20 & 39.959 & 23.539 & 132.6 & 140.4 & 142.0\\
 21 & 39.948 & 25.499 & 143.5 & 143.9 & 145.6 & &  22 & 39.942 & 28.420 & 159.9 & 148.7 & 150.7\\
 23 & 39.943 & 31.330 & 176.3 & 153.2 & 155.4 & &  24 & 39.936 & 34.193 & 192.4 & 157.4 & 159.7\\
 25 & 39.944 & 37.110 & 208.9 & 161.2 & 163.7 & &  26 & 39.960 & 39.968 & 225.1 & 164.8 & 167.5\\
 \end{tabular}
\caption{Results for $\Gamma = 1.003$ from the medium apparatus ($D = 24.84$ cm).}
\label{tab:1.0m}
\end{center}
\end{table}

\begin{table}
\begin{center}
\begin{tabular}{cccccccccccccc}
No& $\bar T (^\circ C)$ & $\Delta T (^\circ C)$ & $10^{-6}R$  & $\cal N$ & ${\cal N}_\infty$ & & No& $\bar T (^\circ C)$ & $\Delta T (^\circ C)$ &  $10^{-8}R$  & $\cal N$ & ${\cal N}_\infty$ \\
 1 & 39.995 &  0.571 & 18.46 & 20.68 & 20.33 & &  2 & 39.995 &  0.721 & 23.34 & 22.13 & 21.76\\
 3 & 39.995 &  0.914 & 29.58 & 23.86 & 23.47 & &  4 & 39.995 &  1.160 & 37.53 & 25.51 & 25.11\\
 5 & 39.995 &  1.473 & 47.64 & 27.35 & 26.94 & &  6 & 39.995 &  1.871 & 60.52 & 29.28 & 28.86\\
 7 & 39.995 &  2.378 & 76.92 & 31.31 & 30.91 & &  8 & 39.995 &  3.025 & 97.85 & 33.57 & 33.13\\
 9 & 39.995 &  3.846 & 124.44 & 35.93 & 35.50 & &  10 & 39.996 &  4.894 & 158.35 & 38.44 & 38.00\\
 11 & 39.996 &  6.229 & 201.52 & 41.17 & 40.71 & &  12 & 39.996 &  7.927 & 256.49 & 44.02 & 43.55\\
 13 & 39.998 & 10.092 & 326.53 & 47.08 & 46.60 & &  14 & 39.999 & 12.848 & 415.73 & 50.35 & 49.87\\
 15 & 39.999 & 16.357 & 529.28 & 53.97 & 53.49 & &  16 & 40.002 & 20.823 & 673.87 & 57.80 & 57.30\\
 17 & 40.035 & 26.550 & 860.19 & 61.92 & 61.41 & &  18 & 40.054 & 33.658 & 1091.23 & 66.23 & 65.71\\
 19 & 40.025 & 33.727 & 1092.34 & 66.37 & 65.85 & &  20 & 40.021 & 35.710 & 1156.37 & 67.41 & 66.90\\
 21 & 40.051 & 37.630 & 1219.84 & 68.45 & 67.93 & &  22 & 40.080 & 39.579 & 1284.33 & 69.41 & 68.90\\
\hline
 1 & 39.996 &  0.636 & 20.58 & 21.23 & 20.88 & &  2 & 39.996 &  1.026 & 33.21 & 24.44 & 24.06\\
 3 & 39.997 &  1.660 & 53.70 & 28.13 & 27.72 & &  4 & 39.999 &  2.691 & 87.08 & 32.23 & 31.81\\
 5 & 40.002 &  4.348 & 140.72 & 37.00 & 36.56 & &  6 & 40.007 &  7.044 & 227.98 & 42.43 & 41.98\\
 7 & 40.008 & 11.433 & 370.05 & 48.69 & 48.19 & &  8 & 40.018 & 18.527 & 599.87 & 55.89 & 55.39\\
 9 & 40.049 & 30.001 & 972.45 & 64.19 & 63.68 & &  10 & 40.065 & 39.567 & 1283.28 & 69.68 & 69.16\\
 \end{tabular}
\caption{Results for $\Gamma = 0.967$ from the small apparatus ($D = 9.21$ cm). Top section: run 1. Bottom section: run 2 after the sample had been taken apart and re-assembled.}
\label{tab:1.0s}
\end{center}
\end{table}

\begin{table}
\begin{center}
\begin{tabular}{cccccccccccccc}
No& $\bar T (^\circ C)$ & $\Delta T (^\circ C)$ & $10^{-8}R$  & $\cal N$ & ${\cal N}_\infty$ & & No& $\bar T (^\circ C)$ & $\Delta T (^\circ C)$ &  $10^{-8}R$  & $\cal N$ & ${\cal N}_\infty$ \\
 1 & 39.901 & 17.562 & 233.67 & 164.7 & 172.6 & &  2 & 40.012 & 15.429 & 206.09 & 158.5 & 165.7\\
 3 & 39.898 & 13.727 & 182.62 & 152.6 & 159.2 & &  4 & 39.986 & 11.633 & 155.24 & 145.1 & 151.0\\
 5 & 39.956 &  9.763 & 130.14 & 137.4 & 142.6 & &  6 & 39.959 &  7.837 & 104.48 & 128.4 & 132.8\\
 7 & 40.089 &  5.663 & 75.85 & 116.2 & 119.7 & &  8 & 39.984 &  3.928 & 52.42 & 103.6 & 106.2\\
 9 & 40.010 &  1.939 & 25.90 & 83.5 & 85.0 & &  10 & 39.970 &  1.041 & 13.88 & 69.3 & 70.2\\
 11 & 39.959 &  3.006 & 40.08 & 95.4 & 97.5 & &  12 & 40.031 &  4.803 & 64.19 & 110.3 & 113.3\\
 13 & 40.252 &  6.302 & 84.89 & 120.4 & 124.2 & &  14 & 39.905 & 17.563 & 233.71 & 164.6 & 172.4\\
 15 & 39.944 &  8.837 & 117.75 & 132.9 & 137.7 & &\\ 
 \hline
 1 & 39.822 & 19.669 & 260.43 & 170.5 & 179.1 & &  2 & 39.827 & 17.730 & 234.80 & 165.3 & 173.3\\
 3 & 40.025 & 13.517 & 180.25 & 152.1 & 158.6 & &  4 & 41.676 & 12.258 & 172.99 & 150.4 & 156.8\\
 5 & 40.083 &  7.623 & 101.86 & 127.5 & 131.8 & &  6 & 39.973 &  5.900 & 78.53 & 117.6 & 121.2\\
 7 & 40.005 &  3.901 & 51.98 & 103.5 & 106.1 & &  8 & 40.008 &  1.948 & 25.96 & 83.6 & 85.1\\
 9 & 40.051 &  2.838 & 37.88 & 93.8 & 95.8 & &\\ 
 \end{tabular}
\caption{Results for $\Gamma = 1.506$ from the large apparatus ($D = 49.7$ cm).   Top section: run 1. Bottom section: run 2 after the sample had been taken apart and re-assembled.}
\label{tab:1.5l}
\end{center}
\end{table}

\begin{table}
\begin{center}
\begin{tabular}{cccccccccccccc}
No& $\bar T (^\circ C)$ & $\Delta T (^\circ C)$ & $10^{-6}R$  & $\cal N$ & ${\cal N}_\infty$ & & No& $\bar T (^\circ C)$ & $\Delta T (^\circ C)$ &  $10^{-8}R$  & $\cal N$ & ${\cal N}_\infty$ \\
 1 & 40.012 &  1.944 & 1097.6 & 63.86 & 65.06 & &  2 & 39.993 &  3.932 & 2218.8 & 79.11 & 81.17\\
 3 & 40.104 &  5.660 & 3206.3 & 88.60 & 91.31 & &  4 & 39.982 &  7.846 & 4426.2 & 97.92 & 101.37\\
 5 & 39.981 &  9.789 & 5521.8 & 104.75 & 108.79 & &  6 & 40.034 & 11.626 & 6570.7 & 110.63 & 115.22\\
 7 & 39.929 & 13.777 & 7757.4 & 116.43 & 121.58 & &  8 & 39.483 & 14.643 & 8116.3 & 117.92 & 123.23\\
 9 & 39.977 & 10.767 & 6072.7 & 107.91 & 112.24 & &  10 & 40.056 &  6.732 & 3807.7 & 93.38 & 96.47\\
 11 & 40.045 &  4.802 & 2715.1 & 84.25 & 86.66 & &  12 & 39.966 &  3.011 & 1697.7 & 72.74 & 74.41\\
 13 & 39.972 &  1.041 & 587.2 & 52.83 & 53.55 & &  14 & 39.961 & 17.593 & 9917.0 & 125.58 & 131.70\\
 \end{tabular}
\caption{Results for $\Gamma = 2.006$ from the large apparatus ($D = 49.7$ cm).}
\label{tab:2.0l}
\end{center}
\end{table}

\begin{table}
\begin{center}
\begin{tabular}{cccccccccccccc}
No& $\bar T (^\circ C)$ & $\Delta T (^\circ C)$ & $10^{-6}R$  & $\cal N$ & ${\cal N}_\infty$ & & No& $\bar T (^\circ C)$ & $\Delta T (^\circ C)$ &  $10^{-8}R$  & $\cal N$ & ${\cal N}_\infty$ \\
 1 & 39.990 & 17.576 & 2937.8 & 85.32 & 89.59 & &  2 & 40.007 & 17.578 & 2939.7 & 85.40 & 89.68\\
 3 & 40.100 & 15.454 & 2592.9 & 82.13 & 86.04 & &  4 & 39.974 & 13.743 & 2295.8 & 79.17 & 82.76\\
 5 & 40.062 & 11.622 & 1947.5 & 75.26 & 78.47 & &  6 & 39.978 &  9.839 & 1643.8 & 71.42 & 74.26\\
 7 & 40.030 &  5.839 & 977.3 & 60.99 & 62.94 & &  8 & 40.002 &  3.928 & 656.9 & 54.02 & 55.48\\
 9 & 40.016 &  1.941 & 324.7 & 43.75 & 44.61 & &  10 & 39.974 &  1.040 & 173.8 & 36.41 & 36.94\\
 11 & 40.063 &  2.830 & 474.2 & 48.99 & 50.14 & &  12 & 40.054 &  4.807 & 805.2 & 57.47 & 59.17\\
 13 & 40.283 &  6.311 & 1065.7 & 62.65 & 64.74 & &  14 & 39.987 &  8.846 & 1478.4 & 69.18 & 71.82\\
 \end{tabular}
\caption{Results for $\Gamma = 3.010$ from the large apparatus ($D = 49.7$ cm).}
\label{tab:3.0l}
\end{center}
\end{table}

\begin{table}
\begin{center}
\begin{tabular}{cccccccccccccc}
No& $\bar T (^\circ C)$ & $\Delta T (^\circ C)$ & $10^{-6}R$  & $\cal N$ & ${\cal N}_\infty$ & & No& $\bar T (^\circ C)$ & $\Delta T (^\circ C)$ &  $10^{-8}R$  & $\cal N$ & ${\cal N}_\infty$ \\
 1 & 40.000 & 19.734 & 412.4 & 46.09 & 48.62 & &  2 & 39.977 & 17.804 & 371.8 & 44.74 & 47.11\\
 3 & 39.402 & 16.991 & 347.7 & 43.83 & 46.09 & &  4 & 40.134 & 13.567 & 284.9 & 41.32 & 43.31\\
 5 & 40.058 & 11.727 & 245.6 & 39.58 & 41.37 & &  6 & 40.054 &  9.781 & 204.8 & 37.51 & 39.10\\
 7 & 40.137 &  7.637 & 160.4 & 34.95 & 36.30 & &  8 & 40.015 &  5.920 & 123.8 & 32.38 & 33.51\\
 9 & 40.032 &  3.909 & 81.8 & 28.73 & 29.57 & &  10 & 40.019 &  1.944 & 40.7 & 23.62 & 24.14\\
 11 & 40.071 &  2.839 & 59.5 & 26.26 & 26.94 & &  12 & 40.085 &  4.791 & 100.4 & 30.49 & 31.47\\
 13 & 40.070 &  6.800 & 142.5 & 33.72 & 34.96 & &  14 & 40.051 &  3.375 & 70.7 & 27.55 & 28.31\\
 \end{tabular}
\caption{Results for $\Gamma = 6.020$ from the large apparatus ($D = 49.7$ cm).}
\label{tab:6.0l}
\end{center}
\end{table}

\begin{figure}
\centerline{\psfig{file=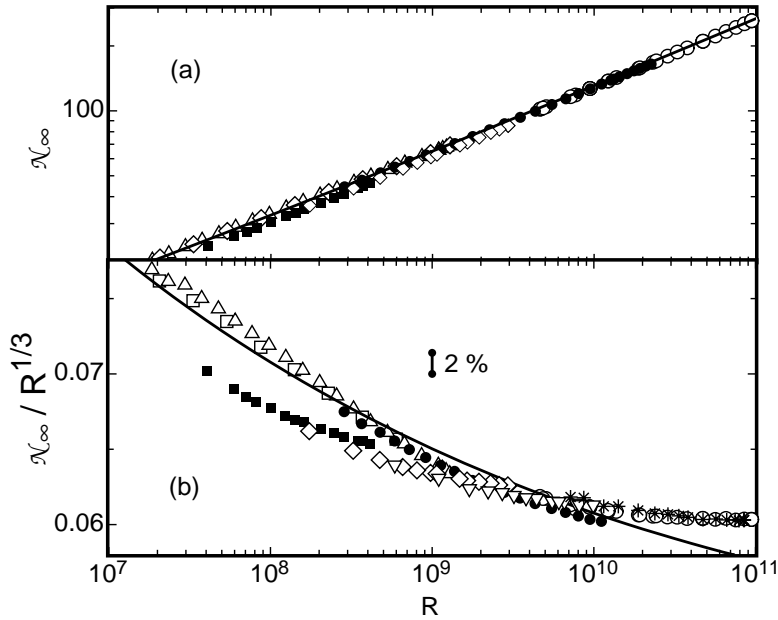,width=4in}}
\caption{(a) The Nusselt number ${\cal N}_\infty$ as a function of the Rayeigh number $R$ on logarithmic scales. (b) The reduced Nusselt number ${\cal N}_\infty/R^{1/3}$ on a linear scale as a function of the Rayeigh number $R$ on a logarithmic scale. Stars: $\Gamma = 0.982,~D = 49.7$ cm, run 1 (from \cite{NBFA05}, corrected for a 0.5\% error in the cross sectional area of the sample).  Open circles: $\Gamma = 0.982,~D = 49.7$ cm, run 2. Solid circles: $\Gamma = 1.003,~D = 24.84$ cm. Open squares (up-pointing triangles): $\Gamma = 0.967,~D = 9.21$ cm, run1 (run2). Open down-pointing triangles: $\Gamma = 2.006,~D = 49.7$ cm. Open diamonds: $\Gamma = 3.010,~D = 49.7$ cm. Solid squares: $\Gamma = 6.020,~D = 49.7$ cm. Solid line: the model of \protect \cite{GL01} for $\Gamma = 1$ and $\sigma = 4.38$.}
\label{fig:nusselt}
\end{figure}

\begin{figure}
\centerline{\psfig{file=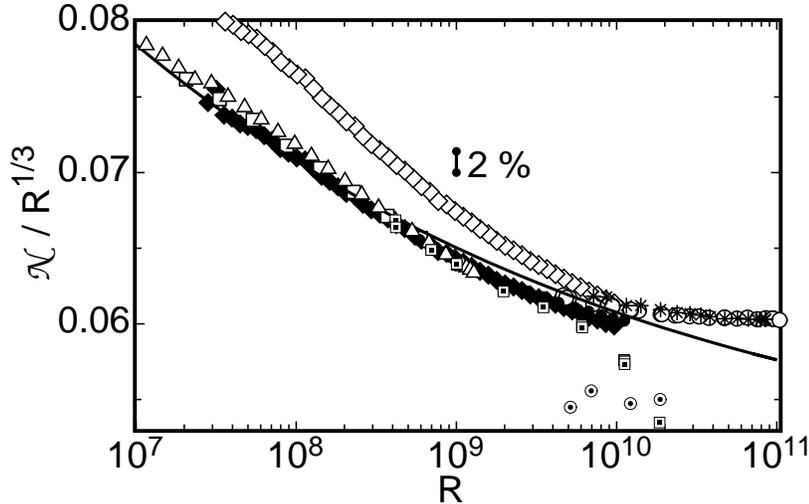,width=4.2in}}
\caption{The reduced Nusselt number  ${\cal N}_\infty/R^{1/3}$ or ${\cal N}/R^{1/3}$ on a linear scale as a function of the Rayeigh number $R$ on a logarithmic scale for $\Gamma \simeq 1$. Stars (open circles): ${\cal N}_\infty/R^{1/3}$ for $\Gamma = 0.982,~D = 49.7$ cm, run 1 (run2). Solid circles: ${\cal N}_\infty/R^{1/3}$ for $\Gamma = 1.003,~D = 24.84$ cm. Open squares (triangles): ${\cal N}_\infty/R^{1/3}$ for $\Gamma = 0.967,~D = 9.21$ cm, run 1 (run2). Open diamonds: ${\cal N}/R^{1/3}$ obtained with acetone 
($\sigma = 3.96$, \cite{XBA00}) for $\Gamma = 1.004$ and $D = 8.74$ cm. Solid diamonds: ${\cal N}_\infty/R^{1/3}$ obtained from the acetone measurements after correction for the wall conductance (\cite{Ah00}). 
Open squares with solid dots: ${\cal N}/R^{1/3}$  obtained by \cite{XLZ02} using water with $\sigma = 4.29$. Open circles with solid dots: ${\cal N}/R^{1/3}$  obtained by \cite{GT79} using water with $\sigma \simeq 6.2$.
Solid line: the model of \protect \cite{GL01} for $\Gamma = 1$ and $\sigma = 4.38$.
}
\label{fig:nussred}
\end{figure}

\begin{figure}
\centerline{\psfig{file=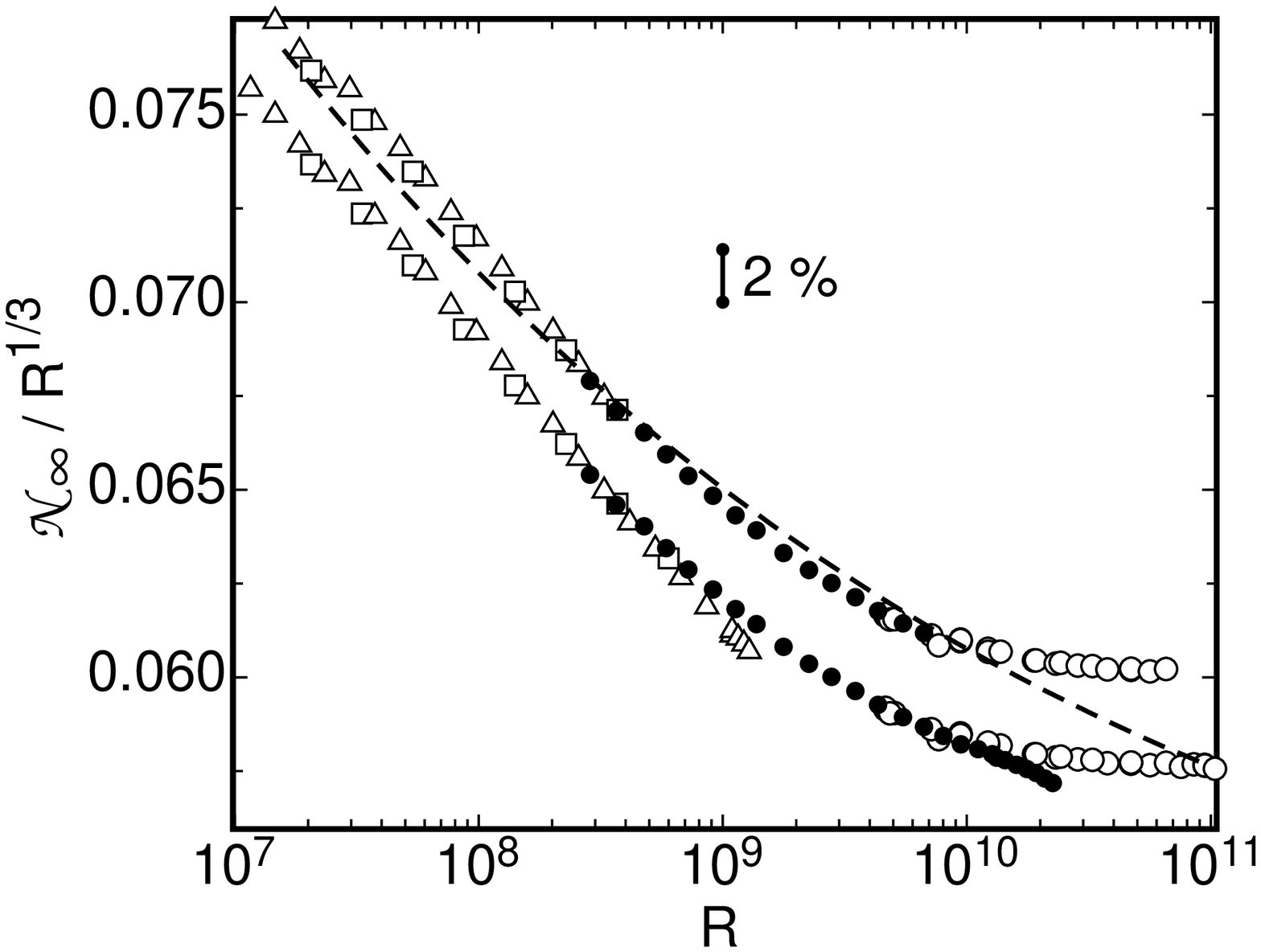,width=4.2in}}
\caption{The reduced Nusselt number ${\cal N}_\infty/R^{1/3}$ on a linear scale as a function of the Rayeigh number $R$ on a logarithmic scale. Open circles: $\Gamma = 0.982,~D = 49.7$ cm, downshifted by 0.3\%. Solid circles: $\Gamma = 1.003,~D = 24.84$ cm, upshifted by 0.6\%. Open triangle: $\Gamma = 0.967,~D = 9.21$ cm, run 1, downshifted by 0.3\%.  Open squares: $\Gamma = 0.967,~D = 9.21$ cm, run 2. Lower set: all data, moved down by 0.0025. Upper set: data that conform ``strictly" to the Boussinesq approximation. Dashed line: the model of \protect \cite{GL01} for $\Gamma = 1$ and $\sigma = 4.38$.}
\label{fig:nussred2}
\end{figure}

\section{Problems associated with high-precision measurements of ${\cal N}$}

One problem  in the measurements of ${\cal N}(R)$ is that data with a precision of 0.1\% or so can be obtained in a given sample only over a range of $R$ covering a little more than a decade unless the fluid is changed. The reason is that the useful temperature differences with conventional fluids like water are limited at the high end to $\Delta T \stackrel {<}{_\sim} 15^\circ$C by possible contributions from non-Boussinesq effects (\cite{Bo1903}) and at the low end  to $\Delta T \stackrel {>}{_\sim} 1^\circ$C by thermometer resolution. For this reason we built three apparatus containing samples of diameter $D = 49.7, 24.8,$ and 9.2 cm, all with $\Gamma \simeq 1$ and known as the large, medium, and small apparatus or sample respectively (\cite{BNFA05}).  Together the data obtained with these span the range $10^7 \stackrel {<}{_\sim} R \stackrel {<}{_\sim} 10^{11}$. 

A second experimental problem is the influence of the side wall on the heat transport by the fluid (\cite{Ah00,RCCHS01,NS03}). Because of the nonlinear temperature profile in the wall adjacent to the thermal boundary layers in the fluid, the heat entering (leaving) the wall at the bottom (top) can be much larger than an estimate based on a constant temperature gradient. In the present work we substantially reduced this problem by choosing a wall of small conductivity (plexiglas or lexan) and a fluid of relatively large conductivity (water). An estimate [model 2 of \cite{Ah00}] indicated that the side-wall corrections for the large and medium samples were less than a few tenths of a percent; they were neglected. For the small sample the correction was 1.7\% for $R = 2\times 10^7$ and smaller at larger $R$, and was made [\cite{BNFA05}] using model 2 of \cite{Ah00}. We believe that for all the data the systematic errors due to the side-wall correction is significantly less than one percent.

A third problem is the effect of the finite conductivity $\lambda_p$ of the confining top and bottom plates on the heat transport by the fluid [\cite{CCC02,Ve04,CRCC04}]. We investigated this effect experimentally [\cite{BNFA05}] by making measurements for samples of different sizes and aspect ratios, each  with copper plates ($\lambda_p = 391$ W/m K) and  with aluminum plates ($\lambda_p = 161$ W/m K). For the large and medium apparatus a small difference between the data sets enabled us to derive a correction factor. When applied to the data taken with the copper plates it yielded an increase of less than 5\% for the large and less than 1\% for the medium apparatus and gave a good estimate of the idealized ${\cal N}_\infty$. For the small apparatus the results obtained with copper and aluminum plates agreed with each other.  

\section{Results}
\label{sec:results}

\subsection{The data}

The measurements were made at a mean temperature of 40$^\circ$C, where $\sigma = 4.38, \kappa = 1.52\times 10^{-7}$ m$^2$/s, $\nu = 6.70\times 10^{-7}$ m$^2$/s, $\alpha = 3.88\times 10^{-4}$ K$^{-1}$, and $\lambda = 0.630$ W/m K. We never observed long transients like those reported by \cite{CRCC04b} for $\Gamma = 0.5$ (see \cite{BNFA05}). On occasion we tilted the apparatus by 2$^\circ$, and within our resolution of 0.1\% found no effect on $\cal N$.  

The results for ${\cal N}$ and  ${\cal N}_\infty$ are given in Tables~\ref{tab:1.0l} to \ref{tab:6.0l} and are shown on logarithmic scales in Fig.~\ref{fig:nusselt}a. With greater resolution they are shown in the compensated form ${\cal N}/R^{1/3}$  in Fig.~\ref{fig:nusselt}b.  The results for $\Gamma = 0.982$ in Table~\ref{tab:1.0l} are not the same as those reported previously (run 1, \cite{NBFA05} Table 4; those results for $\cal N$ and ${\cal N}_\infty$ should be reduced by 0.5\% because of an error in the area used in the original data analysis). They were taken in a second experiment (run 2) after the sample had been taken apart and re-assembled. Likewise, there are two separate runs for $\Gamma = 0.967$ in the small apparatus (Table~\ref{tab:1.0s}) and for $\Gamma = 1.506$ in the large apparatus (Table~\ref{tab:1.5l}). Within a given run the measurements were reproducible within one or two tenths of a percent (see, for instance, points 17 and 24 in Table~\ref{tab:1.0l}). The two runs for $\Gamma = 1.506$ (Table~\ref{tab:1.5l}) agree within their scatter of about 0.1\%.  On the other hand, the two runs with the large apparatus for $\Gamma = 0.982$ (Table~\ref{tab:1.0l} and \cite{NBFA05} Table 4), as well as the two runs from the small apparatus (Table~\ref{tab:1.0s}), differ from each other by a few tenths of a percent, but by no more than expected possible systematic errors.

The results for $\Gamma \simeq 1$ from the small, medium, and large samples fall on nearly continuous smooth curves, but close inspection shows that there are small systematic offsets. The data lie close to the GL model (solid line). It is remarkable that the $\Gamma > 1$ data come so close  to the $\Gamma \simeq 1$ results. For instance, the $\Gamma = 6$ values are only about 4\% below the $\Gamma \simeq 1$ measurements. One assumes that the large-$\Gamma$ sample had a much more complex large-scale-flow structure than the single circulating  roll expected  to exist for $\Gamma = 1$. Apparently this has only a very modest influence on the heat transport. 

In Fig.~\ref{fig:nussred} we compare the present results with previous measurements for $\Gamma \simeq 1$ and $\sigma$ close to 4. Data for $\cal N$ obtained using acetone ($\sigma = 3.96$) are shown as open diamonds [\cite{XBA00}].  The corresponding results obtained after a correction for the side-wall conductance [model 2, \cite{Ah00}] are given as solid diamonds. One sees that in this case the wall correction is quite large, reaching about 8 \% for $R = 10^8$ (no plate correction was required in this case, see \cite{BNFA05}). Nonetheless the corrected data for ${\cal N}_\infty$ are in excellent overall agreement with the present results. The open squares with solid dots at their centers represent the results of \cite{XLZ02} using water with $\sigma = 4.29$. Up to $R \simeq 10^9$ they agree extremely well with the present measurements. For larger $R$ they are slightly lower, presumably because of the influence of the finite plate conductivity. Also shown are data from \cite{GT79}. When corrections for the finite plate-conductivity (which had not been made) and the difference in $\sigma$ are considered, they may be regarded as consistent with the present results.

\begin{table}
\begin{center}
\begin{tabular}{cccccccccc}
$10^{-8}R $ & ${\cal N}_\infty$ & $10^{-8}R$  & ${\cal N}_\infty$ & $10^{-8}R$ & ${\cal N}_\infty$ & $10^{-8}R$ & ${\cal N}_\infty$ & $10^{-8}R$ & ${\cal N}_\infty$ \\
\hline
0.0921 &  16.55 &0.1160 &  17.70 &0.1464 &  18.95 &0.1846 &  20.27 &0.2334 &  21.69 \\
0.2958 &  23.40 &0.3753 &  25.04 &0.4764 &  26.86 &0.6052 &  28.78 &0.7692 &  30.79 \\
0.9785 &  33.04 &1.2440 &  35.39 &1.5830 &  37.86 &2.0150 &  40.59 &2.5650 &  43.42 \\
3.2650 &  46.46 \\
\hline
0.1285 &  18.63 &0.2058 &  21.23 &0.3321 &  24.44 &0.5370 &  28.13 &0.8708 &  32.23 \\
1.4070 &  37.00 &2.2800 &  42.43 &3.7010 &  48.69 \\
\hline
 2.857 &  44.72 & 3.661 &  48.00 & 4.763 &  51.95 & 5.864 &  55.19 & 7.227 &  58.66 \\
 9.119 &  62.88 &11.283 &  66.96 &13.749 &  71.07 &17.749 &  76.66 &22.561 &  82.44 \\
27.906 &  88.01 &34.944 &  94.29 &43.297 & 100.67 &54.777 & 108.30 &66.792 & 115.21 \\
66.792 & 115.21 \\
\hline
 46.36 & 102.79 & 46.42 & 102.90 & 48.42 & 104.11 & 50.55 & 105.63 & 71.55 & 117.77 \\
 76.73 & 120.02 & 94.20 & 128.84 & 94.20 & 128.75 & 94.26 & 128.78 & 94.33 & 128.87 \\
122.18 & 139.94 &122.39 & 139.84 &137.57 & 145.40 &188.56 & 160.89 &192.00 & 161.87 \\
231.37 & 172.01 &242.66 & 174.83 &284.99 & 184.23 &327.32 & 192.89 &376.98 & 201.89 \\
470.65 & 217.34 &471.05 & 217.50 &471.35 & 217.56 &563.10 & 230.58 &655.73 & 242.84 \\
655.73 & 242.84 \\

 \end{tabular}
\caption{Boussinesq results for $\Gamma = 1$. From top to bottom, the sections are for the small sample (run 1), small sample (run 2), medium sample, and large sample (run 2).}
\label{tab:Bouss}
\end{center}
\end{table}

\subsection{Strictly Boussinesq data for $\Gamma \simeq 1$}

The influence of departures from the Oberbeck-Boussinesq approximation (OBA) [\cite{Bo1903}] was considered by various authors. Most recently \cite{NS03} (NS) examined the issue in considerable detail in terms of various fluid properties. Unfortunately at present we have no theoretical criteria to decide whether a given variation over the applied temperature difference of a given property  will affect $\cal N$ significantly. Here we provide some insight into this problem from measurements with samples of different sizes but the same $\Gamma$.
  
Where they overlap, there is a small systematic offset between the $\Gamma \simeq 1$ data from the small sample, run 2 on the one hand and the  medium-sample on the other. A similar offset exists between the data from the medium sample, and the large sample run 2. These offsets are well within possible experimental systematic errors. In order to obtain a single internally consistent data set spanning the entire range $10^7 \stackrel {<}{_\sim} R \stackrel {<}{_\sim} 10^{11}$, we shifted the data for ${\cal N}_\infty$ from the small sample, run 2 downward by 0.3\%. We also shifted the medium-sample data upward by 0.6\%, and those from the large sample, run 2 downward by 0.3\%. The result is shown by the lower sets of data (displaced downward by 0.0025 for clarity) in Fig.~\ref{fig:nussred2}. The results from all three samples now merge smoothly into each other. We can then attribute the deviations of the small-sample data at their largest values of $R$ from the medium-sample data  to deviations from the OBA. A similar situation prevails with respect to the deviations of the medium-sample data from the large-sample results for $R \stackrel {>}{_\sim} 10^{10}$. 

The upper sets of data in Fig.~\ref{fig:nussred2} (plotted without any vertical shift) consist only of those points, taken from the lower sets, that fall within approximately 0.2\% of a smooth, continuous line through all the results. In Table~\ref{tab:Bouss} we give these points in numerical form. We regard these results as conforming ``strictly" to the OBA. They are our best estimate of ${\cal N}_\infty$ for $\sigma = 4.38$ and $10^7 \stackrel {<}{_\sim} R \stackrel {<}{_\sim} 10^{11}$, and constitute the primary result of our work. 

\begin{figure}
\centerline{\psfig{file=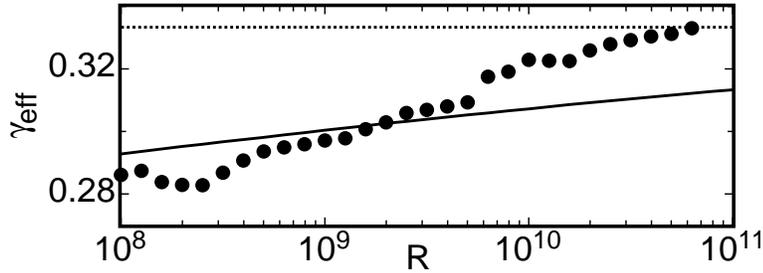,width=4in}}
\caption{Effective exponent $\gamma_{eff}$ of ${\cal N}_\infty$, determined from a powerlaw fit over a sliding window of half a decade in the strictly Boussinesq  range,  as a function of $R$. Dotted line: $\gamma_{eff} = 1/3$. Solid line: result of the GL model.}
\label{fig:gamma}
\end{figure}

\begin{figure}
\centerline{\psfig{file=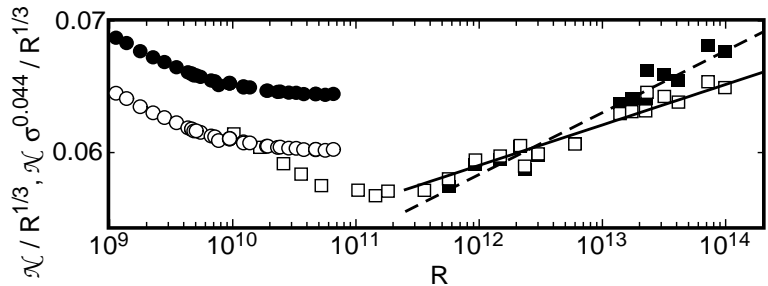,width=4in}}
\caption{${\cal N}/R^{1/3}$ (open symbols) and ${\cal N}\sigma^{0.044}/R^{1/3}$ (solid symbols) as a function of $R$ on logarithmic scales for the present data (circles) and those of \cite{NS03} (squares).}
\label{fig:NScompare}
\end{figure}

\subsection{The effective exponent $\gamma_{eff}$ of ${\cal N}_{\infty}(R)$}

A powerlaw
${\cal N}_\infty = N_0 R ^{\gamma_{eff}}$
was fit to the data for ${\cal N}(R)$ in the strictly Boussinesq range (Table~\ref{tab:Bouss}) within a sliding window covering half a decade of $R$. The results for $\gamma_{eff}$ are shown in Fig.~\ref{fig:gamma}. Near $R = 10^8$  one sees that $\gamma_{eff}$ has a value close to $2/7 \simeq 0.286$, the result of early theories (see, for instance, \cite{Si94}). With increasing $R$ it increases linearly with $log(R)$  within experimental error, reaching  the large-$R$ asymptotic value $\gamma_{eff} = 1/3$ of the GL model at the {\it finite} value $R_0 \simeq 7\times 10^{10}$. Precision measurements conforming to the OBA for $\Gamma = 1, \sigma = 4.4$ and  a wider range of $R$ above $R_0$ are needed to determine whether $\gamma_{eff}$ will remain at 1/3. 

As was seen in Fig.~\ref{fig:nussred2}, the GL model is in reasonable agreement with the experimental results for ${\cal N}(R)$ up to $R \simeq 10^{10}$. However, for the model $\gamma_{eff}$ increases somewhat more slowly with $log(R)$ (solid line in Fig.~\ref{fig:gamma}) and reaches 1/3 only in the limit as $R \rightarrow \infty$ whereas the experimental $\gamma_{eff}$ becomes equal to 1/3 at the finite $R_0 \simeq 7\times 10^{10}$.

The result $\gamma_{eff}\simeq 1/3$ was obtained before by \cite{GT79}. However, they simultaneously fitted all their data,  regardless of $\Gamma$,  over the range $5\times 10^8 \stackrel {<}{_\sim} R \stackrel {<}{_\sim} 3\times10^{11}$ to a power law, and found $\gamma_{eff} \simeq 1/3$ over the entire range. This is not in agreement with our results for $\Gamma = 1$ which yield an $R$-dependent $\gamma_{eff}$. 

An exponent close to 1/3 was found also by NS in experiments for $\Gamma = 1$ using helium gas where $\sigma$ changed with $R$ from about 1 to about 3.8. Those data (open squares) are displayed together with ours (open circles) in Fig.~\ref{fig:NScompare}. Over the range $3\times 10^{11} < R <  10^{14}$ they can be represented by a powerlaw with $\gamma_{eff} = 0.354$ (solid line) (when only data for $R > 10^{13}$ are fitted, one obtains $\gamma_{eff} = 0.345$). The $\sigma$-dependence of $\cal N$ at constant $R$ is not known very well. For $3.62 < \sigma < 5.42$, $\Gamma = 0.67$, and $R \simeq 10^{11}$ we have ${\cal N} \propto \sigma^{-0.044}$ [\cite{NBFA05}]. In order to see how much this could possibly influence the $R$-dependence, we also fitted the NS data for ${\cal N}\sigma^{0.044}$ (solid squares) and obtained $\gamma_{eff} = 0.365$ (dashed line). The results by NS, together with ours, suggest that $\gamma_{eff}$ increases beyond 1/3 as $R$ grows beyond $10^{11}$. 

\section{Acknowledgment}

This work was supported by the US Department of Energy through Grant  DE-FG02-03ER46080.

\end{document}